\documentclass{article}

\usepackage{graphicx}
\usepackage{amsmath}
\usepackage{amssymb}
\usepackage{url}

\begin{document}

\title{On-bird Sound Recordings: Automatic Acoustic Recognition of Activities and Contexts}

\author{Dan Stowell, Emmanouil Benetos and Lisa F. Gill \\ \texttt{\small dan.stowell@qmul.ac.uk}}

\maketitle

\begin{abstract}
We introduce a novel approach to studying animal behaviour and the context in which it occurs,
through the use of microphone backpacks carried on the backs of individual free-flying birds.
These sensors are increasingly used by animal behaviour researchers to study individual vocalisations of freely behaving animals, even in the field.
However such devices may record more than an animal’s vocal behaviour,
and have the potential to be used for investigating specific activities (movement) and context (background) 
within which vocalisations occur.
To facilitate this approach, we investigate the automatic annotation of such recordings through two different sound scene analysis paradigms:
a scene-classification method using feature learning,
and an event-detection method using probabilistic latent component analysis (PLCA).
We analyse recordings made with Eurasian jackdaws (\textit{Corvus monedula}) in both captive and field settings.
Results are comparable with the state of the art in sound scene analysis;
we find that the current recognition quality level enables scalable automatic annotation of audio logger data,
given partial annotation, but also find that individual differences between animals and/or their backpacks limit the generalisation from one individual to another.
we consider the interrelation of `scenes' and `events' in this particular task,
and issues of temporal resolution.	
\end{abstract}


\section{Introduction}
\label{sec:intro}
Studying the behaviour of animals in real time and in their natural environments is becoming more and more feasible through the use of animal-borne loggers or other remote sensing technology \cite{Wilmers:2015}. These technologies have provided insight into different aspects of physiology and behaviour, such as heartbeat \cite{Laske:2011} or migratory routes \cite{Schofield:2010,Newman:2012}, which in turn can help us understand basic mechanisms up to evolutionary drivers, as well as support decision-making processes in nature conservation or disease management.

To reconstruct daily activity patterns, many remote-sensing studies have used methods that provide information on the location of an animal in space (today most commonly GPS: Global Positioning System). To get more fine-scale information, spatial data have been combined with accelerometry  which can shed more light on the actual activities of an animal \cite{Shamoun:2012,Wilmers:2015}. However, the immediate causes or related contexts of specific animal behaviours were often not identifiable through these technologies, and required additional information sources.

Recently, microphone backpacks have become useful tools to investigate different aspects of vocal behaviour in naturalistic contexts, even in small animals \cite{Hiryu:2008,Ilany:2013,Couchoux:2015,Gill:2016}. By picking up the vocal sounds close to their production origin, researchers are now able to record and identify vocalisations from the signal-emitting individuals, even in physically or acoustically challenging environments.
Recording close to the origin also reduces the influence of propagation effects on the audio suchas dispersion or echoes.
But in small animals, unlike for example in whales \cite{Stimpert:2015}, it is often not (yet) possible to apply tags that provide multiple channels of information simultaneously, due to weight limitations---especially in birds. Thus, placing vocal behaviour into relevant context can be limited to specific situations in which a simultaneous collection of further data is possible.

Because an on-board microphone moves along with its bearer, most microphone backpacks do not exclusively record vocalisations, but also other sounds. Firstly, depending on their sensitivity, the microphones have the potential to pick up a variety of background sounds. Secondly, specific movement patterns of the animal resulting in characteristic sound patterns might reveal aspects of the animal's behaviour, e.g. “running” or “self-scratching” (noted by \cite{Ilany:2013,Couchoux:2015}). But, to date, this has not been investigated in detail. 

\subsection*{Automatic Acoustic Recognition}
Successful identification of animal-related sounds could provide a unique opportunity because it may allow investigating not only the behaviour of the animal itself, but also different aspects of its abiotic and biotic environment---which is currently not possible by recording the spatial position or movement of single individuals, without further data collection.
This in turn  could be useful for various purposes (as above: from basic research to conservation, e.g. effects of anthropogenic noise), but analysing such signals/soundscapes remains a challenge to date.
Manual annotation is possible for small datasets, though hard to scale up;
further, for free-flying birds there will usually be no visual/video support for manual annotation. Hence there is strong potential for microphone backpack methodologies to be augmented by automatic acoustic recognition of bird activities and their contexts.

The problem of automatic animal context recognition from audio is directly related to the emerging field of \emph{sound scene analysis} (also termed \emph{acoustic scene analysis}), and more specifically to the two core problems in the field, namely \emph{sound scene analysis} and \emph{sound event detection} \cite{Stowell:2015}. Since the context in question can refer either to an animal's current activity or background sounds, the problem can be viewed as either or both of searching for specific acoustic events (e.g. related to flapping wings in the context of flying) or evaluating the overall properties of a continuous sound scene (e.g. background sounds indicating that an individual is based in a nest).

The vast majority of approaches in the field of sound scene analysis either fall directly into the problem of sound scene recognition (which typically refers to identifying scenes based on location-specific characteristics, e.g. park, car, kitchen) or the problem of sound event detection (which refers to identifying instances of sound events with a start and end time, e.g. door slam, scream) \cite{Stowell:2015}. An approach that is closer to the present work is proposed by Eronen et al. \cite{Eronen06}, who developed a computationally efficient classification-based system for audio-based context recognition in urban environments, where `context' referred to both locations (e.g. train, street) but also to specific activities (e.g. construction, meeting). In \cite{Heittola13}, Heittola et al. proposed a system for sound event detection, which is however dependent on the context of each sound scene. A system based on hidden Markov models (HMMs) with multiple Viterbi decoding was proposed, which was able to identify to a relative degree of success 60 types of sound events, being present in 10 different types of location-dependent audio-related contexts.

Another related strand of research is \textit{speaker diarisation},
in which multi-party speech recordings are analysed such as discussions in meetings,
and the primary goal is to recover a transcript of which party spoke when
\cite{Tranter:2006,Anguera:2012}.
In speaker diarisation, the emphasis is primarily on speech and so the range of sound types considered is often highly constrained.
Also the targets of transcription are individual speaking sources rather than aggregate contextual categories.
Much work in speaker diarisation treats the transcription task as monophonic (only one speaker at a time),
although recent directions are beginning to address overlapping speech \cite{Anguera:2012}.
Generalisation across different domains (e.g.\ conference meetings versus broadcast news) is also an open topic, indicating the difficulty of these types of problem in general.


When placing the present study in context with related work in sound scene analysis, it is important to maintain a focus on the downstream use of the data, which must influence the way we design \textit{and} evaluate systems.
Typical applications in animal behaviour include:
(a) aggregating timelines to produce an overall model of a species' diurnal cycle of activity, or creating ``time budgets'';
(b) data-mining to search for one or many instances of a particular phenomenon.
A transcript is rarely the end goal in itself.
As an example consequence of this, for the applications just mentioned it may often be helpful to obtain a probabilistic or confidence-weighted output rather than merely a list of events, for optimal combination of information or best guidance of subsequent manual effort.

\subsection*{Aims}
The aims of this study were thus to find out whether the recordings from microphone backpacks could be useful for investigating the immediate context in which individual vocalisations occur, such as an animal's current activity (movement sound) or vocalising conspecifics (background sound),
and to investigate the extent to which this could be facilitated by automatic acoustic recognition.
To do so, we used video-validated and human-coded on-bird sound recordings from captive and free-flying jackdaws (\textit{Corvus monedula}), to test the performance of different automatic recognition algorithms.
We experimentally compared two different sound recognition paradigms (classification and event detection),
as well as combinations and variants,
and how they performed in terms of recognising the various categories of activity and context that are of interest for measuring animal behaviour.

In the following we describe the data collection process (Section \ref{sec:collection})
before giving details of our two automatic recognition systems (Section \ref{sec:recognition}).
Our evaluation method and its results are presented in Section \ref{sec:evaluation},
and then in discussion (Section \ref{sec:discussion}) we consider the implications of our study for the automatic annotation of animal-attached sound recordings.

\section{Data Collection}
\label{sec:collection}

\subsection{Birds and microphone backpacks}
For the current study, we used a subset of on-bird sound recordings obtained during a different study (Gill et al., in preparation). The analysed data were collected in the South of Germany, from 12 individual jackdaws (\textit{Corvus monedula}, 7 captive-housed and 6 free-living), early in the years of 2014 and 2015. Backpack application was approved by the Government of Upper Bavaria and in compliance with the European directives for the protection of animals used for scientific purposes (2010/63/EU).
The backpacks consisted of a commercially available digital voice recorder (Edic Mini Tiny A31, TS-Market Ltd., Russia), a rechargeable battery (ICP581323PA to ICP402035, Renata, Switzerland), a radio transmitter for relocation (BD-2 Holohil, Canada) and a shrinking tube casing. Loggers were charged, programmed and read out via PC connection and the according software (RecManager, version 2.11.19, Telesystems, Russia). They were set to record continuously for a few hours every morning, for a few days, beginning one day post capture (at 22050 Hz sampling rate, uncompressed .wav format). This provided coherent vocalisation data and acoustic background information, as opposed to using amplitude-based triggers (but at a cost of storage and battery).
For backpack attachment, birds were either trained to fly inside a smaller compartment of the aviary where they were caught using bird nets (captivity), or trapped inside their nest boxes (wild). The backpacks were fitted using approved attachment methods (glue, or via a harness similar to \cite{Karl:1987}), and following common recommendations ($<5\%$ of body weight \cite{Caccamise:1985}; close to centre of gravity \cite{Vandenabeele:2014}). Birds were individually identified by colour rings. After capture and backpack attachment (20 mins $\pm$ 4.1 SD), they were observed using binoculars and/or radio-telemetry, and all of them were immediately able to fly upon release. For further details on procedures and animal welfare, see Gill et al. (in preparation). 

\subsection{Video-validation of sounds}
For a video-validation of on-bird sound data, video footage was collected from the captive birds during backpack recording hours. For this, an observer sat inside the aviary and video-recorded focal birds using a handheld camcorder (JVC Camcorder Everio GZ-MG77E, Japan). 
All sound files used for video validation were processed, played back, visualised (waveform or spectrograms: FFT window size 512, Hann, 0--10000 Hz viewing range, gain 20--35 dB, range 45 dB) and annotated in Audacity (Version 2.0.5) by LFG.
Corresponding sound and video files were cut to match, and were then played back simultaneously, at normal speed (using Audacity, see above, and using VLC, Version 2.1.5). First, the sounds were annotated step-by-step with the corresponding visual information (see Table \ref{tbl:labels}). If the focal bird was temporarily out of sight, this was labelled as missing data. Secondly, labels were added for acoustically distinct background sounds, such as vocalising jackdaws. Next, the annotation track (labels, start and end points) of each recording was exported as a text file. To balance between fine detail and sufficient sample size, the original labels were used to create slightly broader behavioural and contextual categories (Table \ref{tbl:labels}).

An example clip of annotated data is visualised in Figure \ref{fig:erb}(a).
In Supplementary Information we provide videos showing the studied birds in some example contexts,
along with standard and backpack microphone recordings to illustrate the characteristics of the specific kind of sound recordings dealt with in this work.

\subsection{Annotation of field data}
Having worked with hours of sound and video recordings from jackdaw backpacks, we had learned a good deal about the acoustic representation of behaviours and were able to annotate the sounds in new files in almost as much detail as in combination with the according visual information (at least at the behavioural category level). Thus, the field recording subset was annotated by LFG based on aural and visual inspection of sounds, as learned from the captive dataset and from observations in the field, but also taking into account differences in the sounds due to different materials in the field (e.g. walking on different substrates), as well as different durations (e.g. prolonged flight). Two labels were added that had not been recorded in captivity: copulations; begging chicks inside the nest (Table \ref{tbl:labels}).

\begin{table}
\caption{Labelling scheme for the actions/contexts in our recordings. The ``Category'' column gives the class labels used in the present study, with the other columns indicating the broader or more specific labelling used during manual transcription.}
    \begin{tabular}{lll}
\textbf{Sound type}	&	\textbf{Category}	&	\textbf{Label examples} \\
\hline
\textbf{Movement}	&	Flying	&	Flying \\
	&	Walking	&	Run, walk \\
	&	Looking around	&	Look \\
	&	Manipulation	&	Food, stick \\
	&	Self-maintenance	&	Bill-wipe, preen \\
	&	Small movement	&	Turn \\
	&	Shaking	&	Body, head \\
\textbf{Vocalisations}	&	Focal call	&	Contact call \\
	&	Non-focal call	&	Non-focal call \\
	&	Allofeed vocalisation	&	Allofeed vocalisation \\
	&	Background call	&	Bg mobbing \\
\textbf{Background}	&	Carrion crow	&	Carrion crow \\
	&	Chickens	&	Hen, cock \\
	&	Colony sounds	&	Church bells \\
	&	Noise	&	Traffic noise \\
\textbf{Combination}	&	Allofeeding	&	Allofeeding \\
	&	Copulation	&	Copulation \\
	&	Nest	&	Entering nest \\
\textbf{Other}	&	Antenna	&	Antenna \\
	&	NA	&	Missing video \\
    \end{tabular}
\label{tbl:labels}
\end{table}

\section{Automatic recognition}
\label{sec:recognition}
\begin{figure}[pt]
	\centering
	\includegraphics[width=0.6\linewidth,clip,trim=0mm 00mm 0mm 0mm]{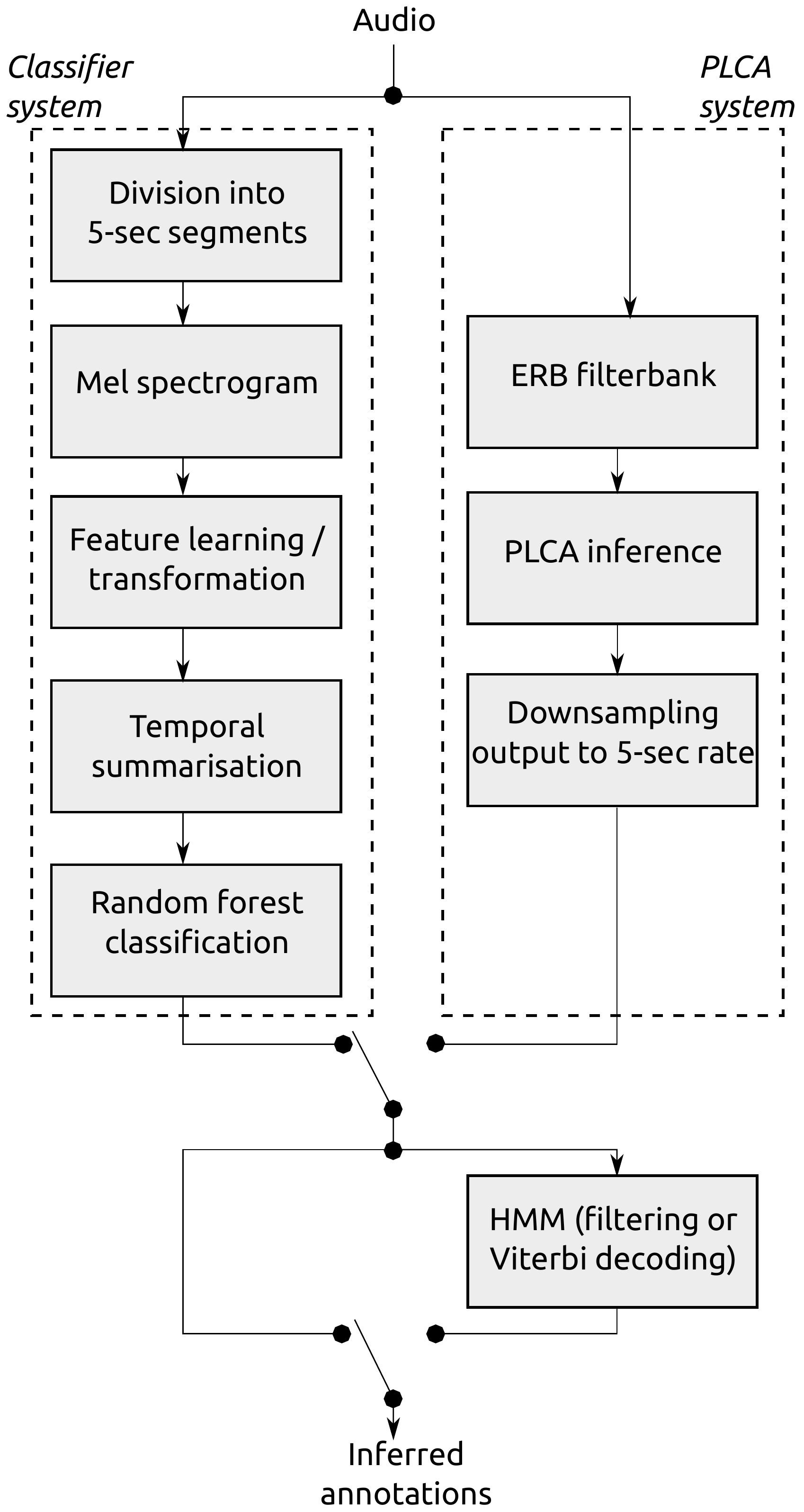}
	\caption{Overview of the processing workflows used for automatic recognition.}
	\label{fig:jcrblockdiagram}
\end{figure}

To train and then test recognition algorithms, we used a total of 8.4 hours of video-validated (captive: 43--100 minutes per bird) and 18.5 hours of human-coded (wild: 164--198 minutes per bird) sound recordings and their respective annotations.
We next describe the automatic recognition systems that we evaluated,
which are summarised in Figure \ref{fig:jcrblockdiagram}.

\subsection{Classifier-based System}

The first system we used for activity and context recognition sits within the classification-based paradigm.
We used our feature learning and classification method previously developed for bird species classification from vocalisations \cite{Stowell:2014b}.
Importantly, this approach applies \textit{spherical k-means} feature learning to Mel-spectrogram patches, in order to transform the input signal into a rich feature space suitable for applying a standard classifier.
This particular feature learning algorithm is conceptually related to an unsupervised convolutional neural network,
but its simplicity makes it eminently scalable to big data \cite{Coates:2012,Stowell:2014b}.
In this work, we segmented input audio into contiguous five-second clips,
from which we calculated Mel spectrograms (FFT window size 1024 with 50\% overlap),
and applied median-clipping noise reduction to each frequency band.
Unlike in the cited previous work, for these data we did not apply high-pass filtering,
since we expected some classes to be indicated in part by lower-frequency or broadband components.
During training we applied a single pass of the feature learning decribed in \cite{Stowell:2014b} to these data,
learning a high-dimensional projection onto 500 features.
We then transformed the training and test data into this new feature space,
before summarising each audio clip by the mean and standard deviation of each feature (i.e.\ 1000 summary features).


The summary features were used as input to a random forest classifier \cite{Breiman:2001} having 200 trees and trained using an entropy-based criterion for splitting branches.
These settings led to good performance in previous work \cite{Stowell:2014b}.
The data in this task is highly unbalanced, with some classes very sparsely represented.
A random forest classifier is typically able to handle unbalanced (and high-dimensional) data well.
However, an option available to us was to reweight the data to give equal prominence to positive and negative classes.
This was particularly pertinent as the subsequent HMM postprocessing (see subsection \ref{sec:hmm}) also makes use of the relative class balance.
We therefore trained the classifier in both modes, equally weighted and balanced-reweighted,
to inspect the effect of this choice.

\subsection{Event Detection System}

The second system used for activity and context recognition is adapted from the system of \cite{Benetos16}, which was originally proposed for sound event detection in office environments. Thus, this approach attempts to recognize contexts as a collection of acoustic events related to each context, as opposed to the previous approach which was based on modelling the overall characteristics of an acoustic scene. The system extends probabilistic latent component analysis (PLCA) \cite{Shashanka08}, a spectrogram factorisation technique which can be viewed as the probabilistic counterpart of non-negative matrix factorization (NMF) \cite{Li99}. The PLCA-based model assumes that an audio spectrogram can be decomposed as a series of sound activities or contexts, which can potentially overlap over time. Each activity is produced as a combination of sound exemplars, which have been pre-computed from training data.

\begin{figure*}
\centering
\resizebox{\linewidth}{!}{
\includegraphics[]{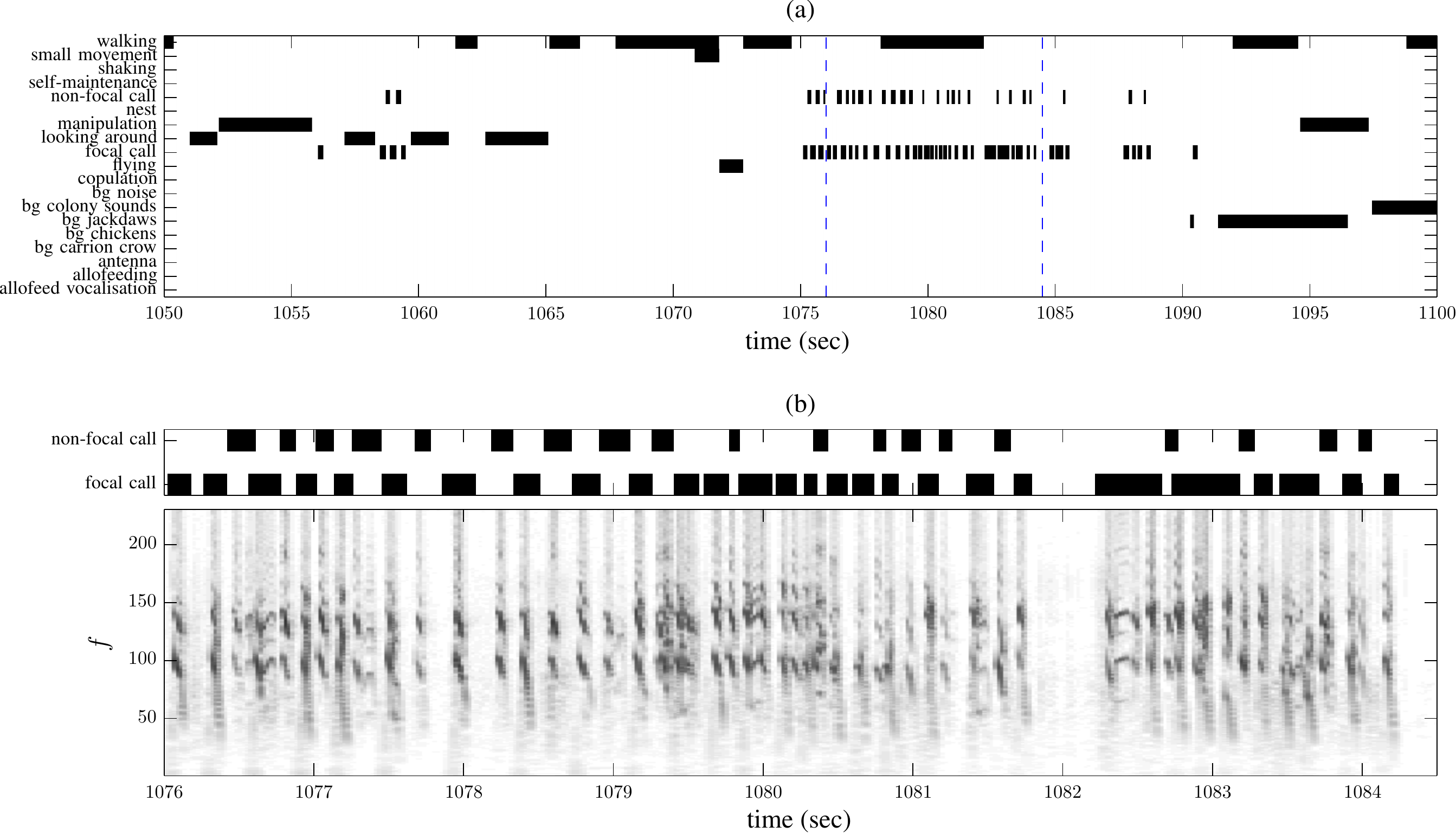}} 
\caption{(a) Context annotations for a recording segment from a captive bird. (b) The annotations for focal and non-focal calls and respective ERB spectrogram of the same recording, both corresponding to the temporal region marked with vertical dashed lines in figure (a).}
\label{fig:erb}
\end{figure*}

For \emph{preprocessing}, a time-frequency representation $V_{f,t}$ ($f$ is the frequency index and $t$ is the time index) is computed by processing the input waveform with an equivalent rectangular bandwidth (ERB) filterbank \cite{Moore95}, using the approach of \cite{Vincent10}. The filterbank uses 250 filters which are linearly spaced between 5~Hz and 10.8~kHz on the ERB scale, and has a 23ms time step. Given that in the context of on-bird sound recordings several activities exhibit information in higher frequencies, a linear pre-emphasis filter is applied to $V_{f,t}$ for boosting high frequency content.
See Figure \ref{fig:erb}(b) for an ERB spectrogram of a recording from the captive subset, along with the respective context annotation.

The PLCA-based model takes as input $V_{f,t}$ and approximates it as a bivariate probability distribution $P(f,t)$, which is in turn decomposed into a series of spectral templates per sound activity/context and exemplar index, activations over time for each context class, as well as an auxiliary probability for the activation of each exemplar per context class over time. The model is formulated as:
\begin{equation}
 P(f,t) = P(t)\sum_{c,e}P(f|c,e)P(c|t)P(e|c,t) \label{eq:plca-model}
\end{equation}
where $c\in\{1,\ldots,C\}$ denotes the context class and $e\in\{1,\ldots,E\}$ denotes the exemplar index. On model parameters, $P(t)=\sum_{f}V_{f,t}$, which is a known quantity. Dictionary $P(f|c,e)$, which in this system is pre-computed from training data, contains spectral templates per context class $c$ and exemplar $e$. The main output of the PLCA model is $P(c|t)$, which is the probability of an active context per time frame $t$. Finally, the model also contains the auxiliary probability $P(e|c,t)$, which denotes the contribution of each exemplar $e$ for producing a context $c$ at time $t$.

The unknown model parameters $P(c|t)$ and $P(e|c,t)$ can be iteratively estimated using the expectation-maximization (EM) algorithm \cite{Dempster77}. For the \emph{E-step}, the following posterior is computed:
\begin{equation}
 P(c,e|f,t) = \frac{P(f|c,e)P(c|t)P(e|c,t)}{\sum_{c,e}P(f|c,e)P(c|t)P(e|c,t)} \label{eq:estep}
\end{equation}
Using the above posterior, $P(c|t)$ and $P(e|c,t)$ can be estimated in the \emph{M-step} as follows:
\begin{equation}
 P(c|t) = \frac{\sum_{e,f}P(c,e|f,t)V_{f,t}}{\sum_{c,e,f}P(c,e|f,t)V_{f,t}} \label{eq:mstep1}
\end{equation}
\begin{equation}
 P(e|c,t) = \frac{\sum_{f}P(c,e|f,t)V_{f,t}}{\sum_{e,f}P(c,e|f,t)V_{f,t}} \label{eq:mstep2}
\end{equation}
Parameters $P(c|t)$ and $P(e|c,t)$ are initialised in the EM updates with random values between 0 and 1 and are normalised accordingly. Eqs. (\ref{eq:estep}) and (\ref{eq:mstep1})-(\ref{eq:mstep2}) are iterated until convergence. In our experiments, we found 30 iterations to be sufficient. 

In order to extract dictionary $P(f|c,e)$ from training data, first spectra $V^{(c)} \in \mathbb{R}^{F\times T_{c}}$ that correspond to an active context class are collected, where $T_{c}$ corresponds to the number of spectral frames that contain an active context class $c$. Then, for each context class a list of exemplars is created by performing clustering on $V^{(c)}$ using the k-means algorithm; here, the number of exemplars $E=40$, following experiments on the training data.

The output of the PLCA model is given by $P(c,t) = P(t)P(c|t)$, i.e. the context activation probability, weighted by the energy of the spectrogram. Since $P(c,t)$ is a non-binary representation, it needs to be converted into a list of estimated contexts per time frame. The first option of post-processing $P(c,t)$ is by performing thresholding, where threshold values were estimated per context class using training data. Finally, active contexts with a small duration (shorter than 120ms) were removed. Additional post-processing options are discussed in the following subsection.

\subsection{Postprocessing} \label{sec:hmm}

Given the output from either the classifier or PLCA detector, we then optionally applied hidden Markov model (HMM) postprocessing to the estimated event sequences. See \cite{Rabiner89} for an overview of HMMs.
HMM-based postprocessing is a common procedure using knowledge about the temporal structure of event sequences (gleaned from the training set) which knowledge may not otherwise be reflected.
In particular, in our case the classifier treats each five-second segment as independent, neglecting information from neighbouring segments. Likewise, the PLCA event detection system considers each 23~msec output frame as independent.

Since our task was polyphonic, having multiple ``channels'' in parallel whose activation could be on or off, there was a combinatorially large number of possible states at any time ($2^k$, with $k$ the number of classes).
To deal with this large state space we applied the HMM in two alternative ways:
(a) applying a single HMM to the entire system, whose set of possible states is the whole set of state combinations observed in the training data; 
or
(b) independently applying a two-state, on/off HMM to the data of each class.
Each approach has advantages and drawbacks.
Treating channels as independent may lead to efficient training given a limited amount of data,
but it neglects interaction effects which could help to resolve ambiguous situations.
Therefore we tested both approaches.

We trained the HMMs generatively, using Laplacian smoothing of the transition tables---i.e.\ initialising each possible transition with a small uniform weight,
which yields a prior equivalent to having observed one instance of each possible transition.
The emission model for each HMM state was a Gaussian mixture model (GMM).
To initialise and to select the number of GMM components, we applied the \textit{Dirichlet process GMM} approach \cite{Rasmussen:1999} to the entire training dataset (sometimes called a \textit{universal background model} or UBM), then for each HMM state we trained its emission model by variational inference initialised from the UBM.
We used the GMM implementations provided by \textit{scikit-learn} 0.17 \cite{Pedregosa:2011}.

Having trained a HMM, there are multiple ways to apply it to new data.
We explored the use of forward filtering---producing probabilistic ``fuzzy'' output which may then be thresholded if definite decisions are required---and Viterbi decoding---producing a single definite output, as the maximum likelihood state sequence given the observations.
This then resulted in four kinds of HMM postprocessing:
filtered or Viterbi-decoded output, from a jointly or independently-trained HMM.

\subsection{Handling Missing Data}
Occasional time-regions of the data were labelled as missing data (`NA'), when birds were occasionally off-camera.
These regions (around 17 minutes total, out of the 8.4 hours of captive audio) were excluded from the training of the classifiers and HMMs.
For the PLCA-based system, the NA class was not used to create the pre-extracted dictionary $P(f|c,e)$, and any spectral frames belonging to the NA class were not used in the training data.
In the test phase, any NA regions in the ground truth are set to be non-active, where any time frames $t$ in the PLCA model output that correspond to the NA regions are set so that $P(c,t)=0$.
`NA' regions were excluded from the calculation of our evaluation statistics,
due to the lack of ground truth for comparison.


\section{Evaluation}
\label{sec:evaluation}
\subsection{Metrics}

As discussed in Section \ref{sec:intro}, the evaluation must be designed with regard to the planned or typical downstream use case---i.e.\ what tasks or analyses do we expect to follow on from such automatic annotation?
For the present task, this bears upon the figures of merit which one calculates, as well as on issues such as the temporal granularity or temporal tolerance.
It is desirable for an automatic system to recover exactly-timed transcriptions of every vocalisation, action and context given in the audio,
but for some of the downstream tasks we consider the overriding aim does not require the highest resolution, for example when characterising time budgets across large datasets, or locating examples of certain activity.
Hence our main evaluation measures were calculated at a five-second granularity (the same granularity as was used for the classifier).
The output of the classifier-based system was itself at a five-second granularity;
for the PLCA-based system, the output was sampled at 23ms steps, as in the input time-frequency representation $V_{ft}$.
We therefore grouped its outputs into five-second segments, and the output for each 5-sec segment was either the mean or the maximum of the 23~msec-step frames corresponding to that time segment.

Evaluation metrics for automatic transcription have been debated in music informatics and in sound scene analysis.
Recently Mesaros et al.\ reviewed such measures for general sound event detection, discussing issues including the use of high-resolution versus segment-based metrics \cite{Mesaros:2016}.
In their terminology our main metrics are segment-based, using five-second segments.
However, Mesaros et al.\ consider only the evaluation of ``definite'' transcripts,
not transcripts with probabilistic/ranked/fuzzy annotations,
and as a result their review does not include statistics useful for evaluating the latter type of output.
Foster et al., working with probabilistic outputs, use a four-second segment size and use the area under the ROC curve (``AUC'') as their figure of merit \cite{Foster:2015}.

The AUC is widely used as an evaluation measure for detection and classification tasks,
and has many desirable properties \cite{Fawcett:2006}:
unlike raw accuracy, it is not impeded by ``unbalanced'' datasets having an uneven mixture of true-positive and true-negative examples; and it has a standard probabilistic interpretation, in that the AUC statistic tells us the probability that the algorithm will rank a random positive instance higher than a random negative instance.
This last feature makes it particularly suitable to evaluating with regard to downstream tasks in which the subsequent postprocessing will for example involve manually confirming/refining the separation of positive and negative instances.
Hand criticises the AUC statistic \cite{Hand:2013}, but reluctantly confirms that its use is well-founded when the downstream makes use of the ranking information, for example to allocate a fixed budget of manual postprocessing time.

An alternative widely-used evaluation measure is the ``F~score'':
the harmonic mean of precision  (robustness against false positives) and recall (robustness against false negatives) of a system \cite{Mesaros:2016}.
The F score is particularly suited to information-retrieval type applications, such as downstream tasks in which the user might for example wish to retrieve a subset of positive examples from a large database.
The F score requires definite, binarised output;
for fuzzy outputs, this requires postprocessing such as thresholding.

In the present work we calculated both the AUCs and the F scores for our systems,
yielding slightly different perspectives on their relative performance.
Both measures were calculated from the segment-wise output with five-second segment durations.
AUCs were calculated separately for each class (our plots will show averages across classes).
To use the F score with fuzzy outputs, we chose binarisation thresholds to optimise the score on the training data, before applying the same thresholds to the testing data in each case.
This can be done with one threshold per class or with a single threshold;
we tested both variants.
To summarise the F score we calculated it across all classes, rather than averaging the per-class F scores, since the latter would be numerically unstable especially with sparse data \cite{Mesaros:2016}.

\subsection{Evaluation Schemes}

Our data consisted of annotated long-duration audio from multiple individual birds, one set in captive conditions and one set in field conditions, with multiple recordings from each individual (3--8 per individual for captive; 2 per individual for field, of longer duration).
We used this data to evaluate system performance in various crossvalidation scenarios:
\begin{description}
\item[\textit{EachCap}: Captive, strictly per-individual.] \hfill \\ A system was trained with one half of an individual's recordings, and tested with the other. The converse was also done, and then results aggregated over all captive individuals (yielding 14 `folds').
\item[\textit{X-Y}: Captive, pooled.] \hfill \\ A system was trained with examples from each individual---half of the recordings from each individual---and tested with the remainder. This gave 2 crossvalidation folds. Note that X-Y is constructed so that all the testing files come from birds also seen in the training data.
\item[\textit{A-B}: Captive, pooled and stratified.] \hfill \\ All recordings from each individual were allocated to one of two partitions. This is similar to X-Y except that no bird used for training is used for testing.
\item[\textit{Cap-Field}: pooled cross-condition.] \hfill \\ In this case the captive data is used for training, and the field data used for testing. (Here we used only one crossvalidation fold.) It is the most challenging case: as well as the train and test sets having no birds in common, the recording situation is also different.
\item[\textit{EachField}: Field, strictly per-individual.] \hfill \\ As \textit{EachCap}, but for the field data (12 folds).
\end{description}
Each of these scenarios relates not just to different degrees of generalisation,
but to different downstream applications of automatic recognition technology.
For example, a researcher may wish to annotate a fraction of a recording and then invoke automatic recognition for the remainder;
or to use a fixed system trained on one set of birds, e.g.\ observed in captivity, and to apply it to new unknown recordings.

Finally, since the PLCA-based system produced its output at a higher resolution (i.e.\ for each 23ms frame),
we used this opportunity to explore how the temporal resolution interacts with evaluation procedures and metrics. 
For this we repeated our evaluation using the segment-based F score,
but using a much smaller segment size of 0.1~seconds, as compared with the 5~sec segment size used in the main experiments.
In order to ensure a fair comparison, sets of class-specific thresholds were computed from training data for each evaluation segment size (i.e. 100~msec and 5~sec) separately. The F-measure was computed directly on the raw high-resolution output of the PLCA-based system.

\subsection{Results}
\label{sec:results}
\begin{figure}[t]
	\centering
	{\includegraphics[page=3,height=0.75\linewidth,clip,trim=2mm 0mm 0mm 0mm]{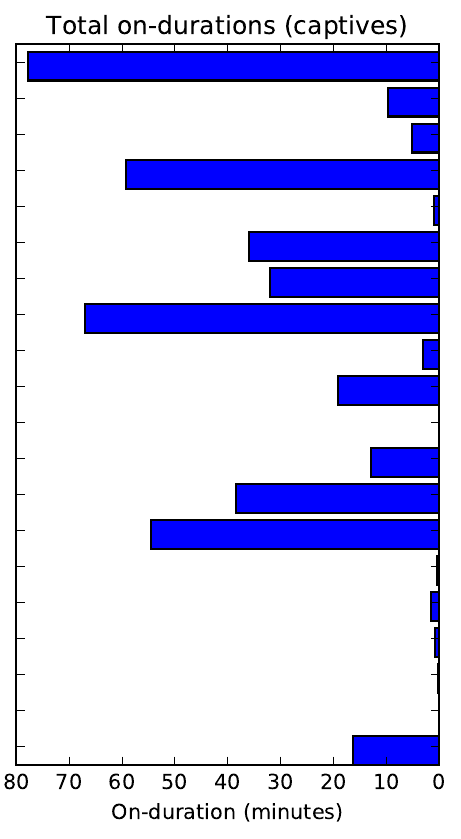}}%
	{\includegraphics[page=3,height=0.75\linewidth,clip,trim=2mm 0mm 0mm 0mm]{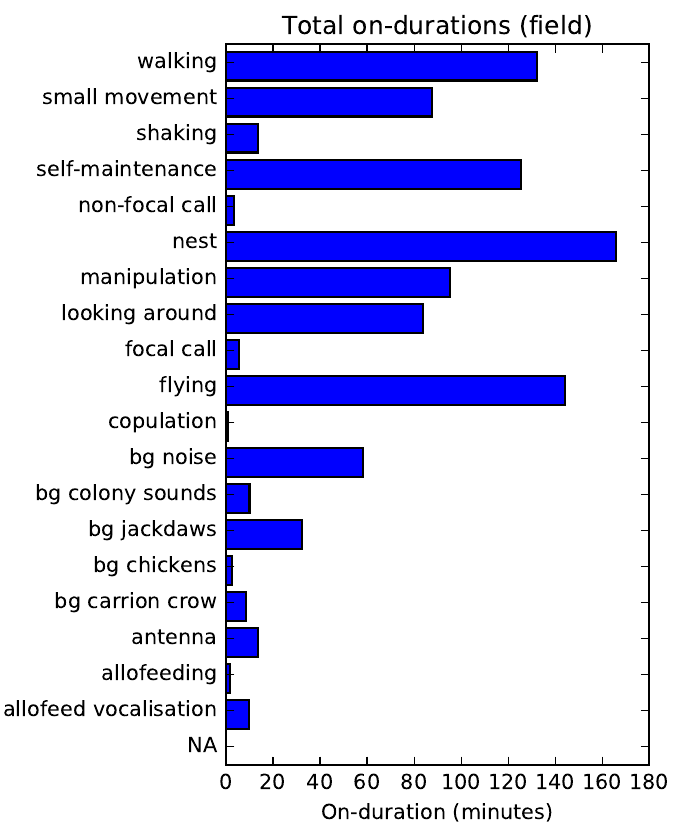}}
	\caption{Total ground-truth durations of annotated regions of each category.}
	\label{fig:ondurations}
\end{figure}

%
As intended, the choice of microphone placement led to high-amplitude recordings for sounds from the focal bird (calls, flying, and other movements) while other background sounds were quiet but still largely audible (see Supplementary Information for examples).
The occurrence of the annotated actions and contexts in the collected data was relatively sparse
(Figure \ref{fig:ondurations}),
with every class being active for less than 16\% of the total time in both datasets.

We evaluated each of our systems in two configurations:
the classifier-based system with unbalanced or balanced class-weighting for training;
and the PLCA system with mean- or maximum-based temporal downsampling.
In each case the differences between configurations were small,
and so for clarity of presentation we will plot results from just one of each system (unbalanced classifier, mean-downsampling PLCA).
We will refer to differences in outcomes from the system configurations where relevant.

\begin{figure*}[pt]
	\centering
	{\includegraphics[page=8,height=0.265\textheight,clip,trim=0mm 20mm 2mm 0mm]{figures/allrunstats}}%
	{\includegraphics[page=10,height=0.265\textheight,clip,trim=10mm 20mm 2mm 0mm]{figures/allrunstats}}%
	{\includegraphics[page=6,height=0.265\textheight,clip,trim=10mm 20mm 2mm 0mm]{figures/allrunstats}}%
	{\includegraphics[page=7,height=0.265\textheight,clip,trim=10mm 20mm 2mm 0mm]{figures/allrunstats}}%
	{\includegraphics[page=9,height=0.265\textheight,clip,trim=10mm 20mm 2mm 0mm]{figures/allrunstats}}%
	\\
	{\includegraphics[page=3,height=0.322\textheight,clip,trim=0mm 0mm 2mm 5mm]{figures/allrunstats}}%
	{\includegraphics[page=5,height=0.322\textheight,clip,trim=10mm 0mm 2mm 5mm]{figures/allrunstats}}%
	{\includegraphics[page=1,height=0.322\textheight,clip,trim=10mm 0mm 2mm 5mm]{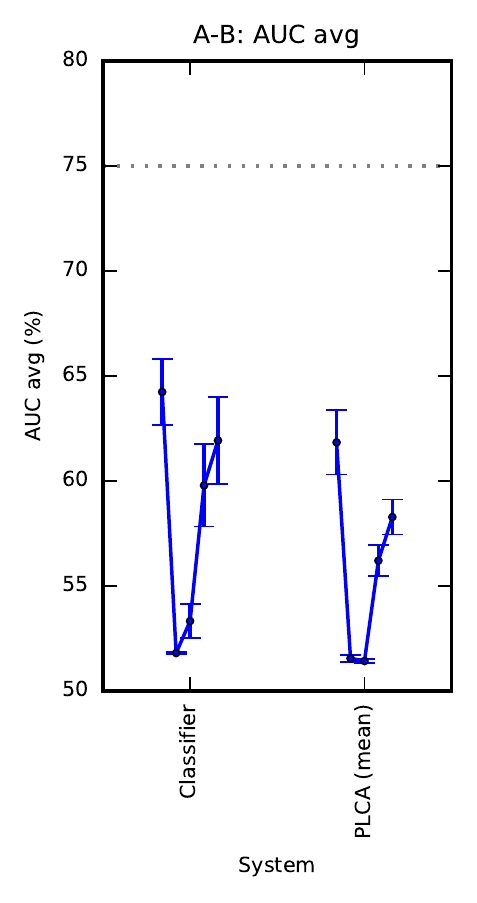}}%
	{\includegraphics[page=2,height=0.322\textheight,clip,trim=10mm 0mm 2mm 5mm]{figures/allrunstats}}%
	{\includegraphics[page=4,height=0.322\textheight,clip,trim=10mm 0mm 2mm 5mm]{figures/allrunstats}}%
	\caption{F scores (top row) and AUCs (bottom row) for the systems tested. Each panel shows a different crossvalidation setup. In each panel, we show clusters of scores connected by lines; the items in each cluster relate to the different postprocessing options, left-to-right as follows: no postprocessing; unified HMM Viterbi decoding; per-class Viterbi decoding; unified HMM filtering; per-class HMM filtering. Plotted values are the median across crossvalidation folds, with error bars indicating their 5- and 95-percentiles.}
	\label{fig:allrunstatsf}
\end{figure*}

Overall, the quality of automatic recognition showed a strong dependency on the choice of crossvalidation setup, i.e.\ on the relationship between the training data and the test data (Figure \ref{fig:allrunstatsf}).
As one clear example: the designs of the X-Y and A-B schemes were very similar except that the latter ensured that birds used for testing were not used for training;
this change incurred a substantial penalty both in AUC and F score, implying that individual differences were highly pertinent.
The X-Y scheme in turn was similar to the EachCap scheme except that it pooled the training data across individuals.
Curiously, this pooling led to very similar F scores as EachCap,
but to a marked difference in AUC:
judged by AUC, the pooling of training data seems to have led to better generalisation properties, for both of the recognition algorithms tested.
Judged by F score, both EachCap and EachField, using systems trained specifically for each individual,
attained many of the strongest results.
As expected, schemes involving generalising to unseen conditions had lower recognition scores---both A-B (generalising to new birds) and Cap-Field (generalising to new birds and to new recording environments).

\begin{figure*}[t]
	\centering
	\includegraphics[width=0.95\linewidth,clip,trim=0mm 6mm 0mm 7.55mm]{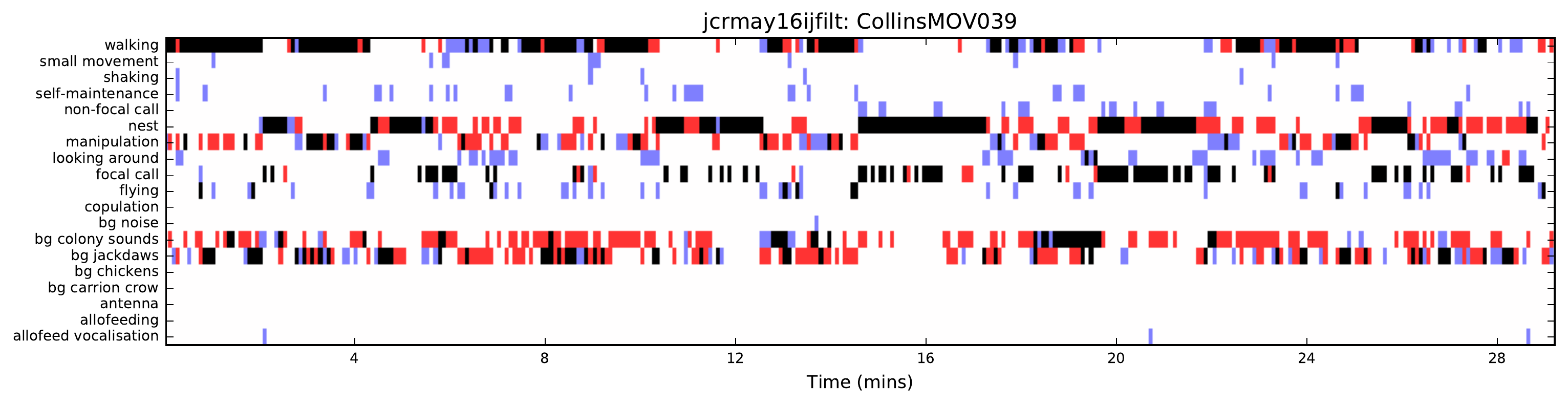}
	\includegraphics[width=0.95\linewidth,clip,trim=0mm 3mm 0mm 7.55mm]{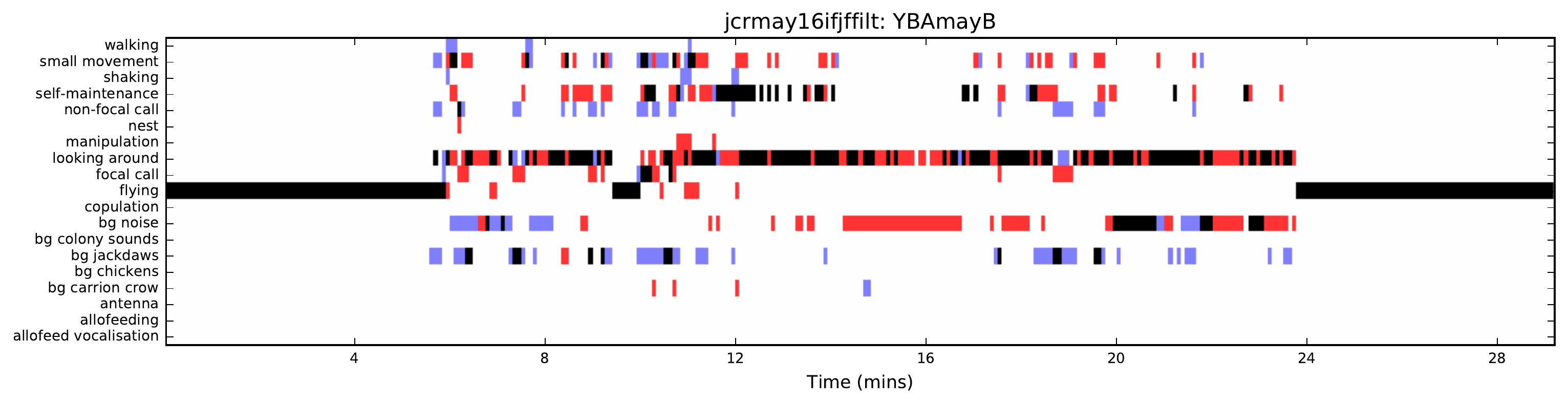}
	\caption{Two examples of automatic annotation from a relatively strongly-performing system (classifier; HMM filtering; per-individual training) for a captive (upper panel) and a field condition (lower panel). The black and white regions are correctly-identified as on and off respectively. Red are false-positive detections, and blue false-negatives. (Best viewed in colour.)}
	\label{fig:wavann1}
\end{figure*}

As this task has not been evaluated before, there are no direct external comparisons for the overall recognition quality.
The segment-wise F-measures are broadly comparable to those presented in \cite{Mesaros:2016} (for an indoor event-detection task with fewer categories and a different segment duration).
In the present comparison of two different approaches,
the classifier-based system generally outperformed the PLCA-based system:
by an average of 5 percentage points on AUC,
and 8 percentage points on F score.
Figure \ref{fig:wavann1} shows an example of the output from the classifier-based system overlaid with the groundtruth annotation,
giving a rough visual indication of the kind of output that corresponds to the results obtained.

The effect of HMM postprocessing led to different results when considered via F score or AUC.
The F score statistics (Figure \ref{fig:allrunstatsf}, upper) often showed a mild improvement when HMM postprocessing is added,
particularly for the classifier-based system;
while the AUC statistics (Figure \ref{fig:allrunstatsf}, lower) unanimously indicated worse results with HMM postprocessing (the leftmost result in each cluster, the unprocessed output, performing best).

To binarise continuous-valued output, we found that per-class thresholding was not particularly better than a single threshold in general, except in the case of the raw PLCA output.
This exception is because the raw PLCA output is expressed in terms of activation magnitude (i.e. related to the energy of each context class in the spectrogram), which does not have comparable meaning across classes, and so per-class thresholding is highly pertinent in that case.
For the HMM-postprocessed outputs, a single threshold often slightly outperformed per-class thresholds, which is probably due to a slight reduction in overfitting the threshold choice.

\begin{figure*}[pt]
	\centering
	{\includegraphics[page=4,height=0.35\textheight,clip,trim=0mm 0mm 2mm 0mm]{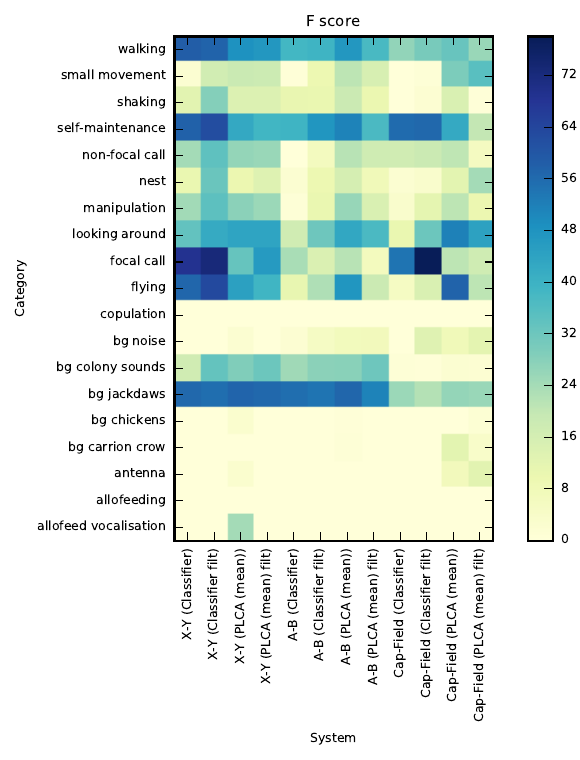}}%
	\hspace{5mm}
	{\includegraphics[page=5,height=0.35\textheight,clip,trim=29mm 0mm 2mm 0mm]{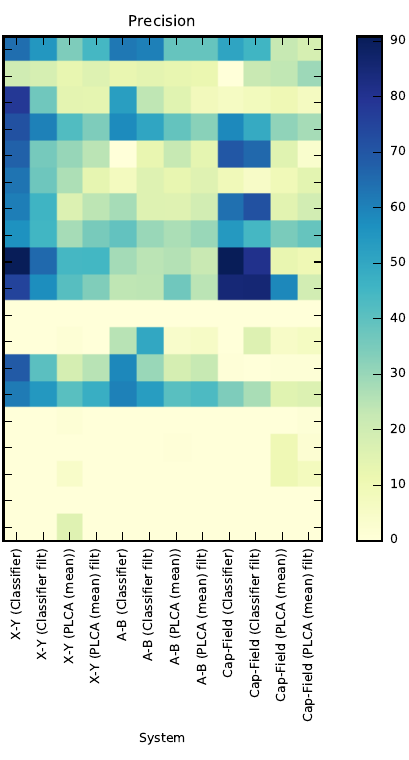}}%
	\hspace{5mm}
	{\includegraphics[page=6,height=0.35\textheight,clip,trim=29mm 0mm 2mm 0mm]{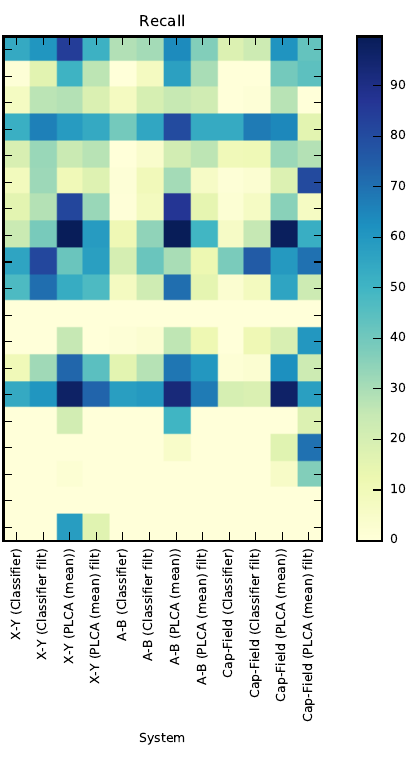}}%
	\caption{F score, Precision and Recall (all in \%) for each class separately, for 4 systems tested under the three pooled crossvalidation scenarios (X-Y, A-B, and Cap-Field), using per-class thresholding.}
	\label{fig:allrunstatsperclass}
\end{figure*}

\begin{figure*}[pt]
	\centering
	{\includegraphics[page=4,height=0.41\textheight,clip,trim=0mm 0mm 2mm 0mm]{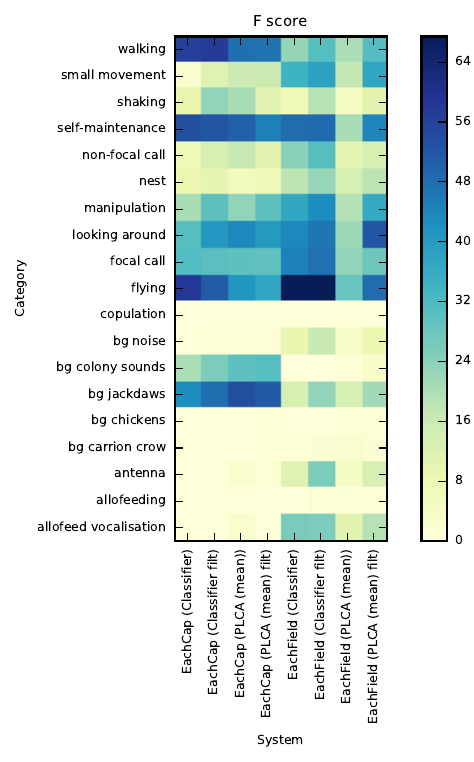}}%
	\hspace{10mm}
	{\includegraphics[page=5,height=0.41\textheight,clip,trim=29mm 0mm 2mm 0mm]{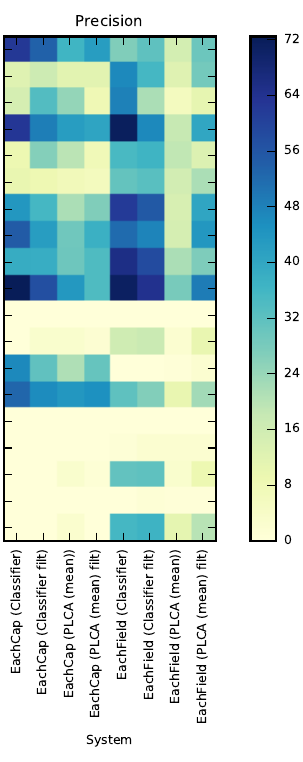}}%
	\hspace{10mm}
	{\includegraphics[page=6,height=0.41\textheight,clip,trim=29mm 0mm 2mm 0mm]{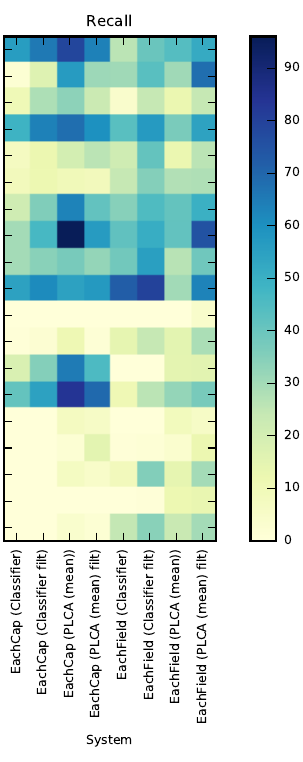}}%
	\caption{Per-class results as in Figure \ref{fig:allrunstatsperclass} but for the two per-individual scenarios (EachCap and EachField).}
	\label{fig:allrunstatsperclass_ind}
\end{figure*}

The classes (categories) used in this study are highly diverse in kind,
and so to drill further into system performance it is important to inspect performance on a per-class level
(Figures \ref{fig:allrunstatsperclass} and \ref{fig:allrunstatsperclass_ind}).
It is immediately clear that detection quality exhibits some correlation with the quantity of positive examples available for training
(cf.\ Figure \ref{fig:ondurations}),
although the \textit{focal call} category is particularly well detected by the classifier system despite being relatively sparse in the training data.
Focal calls are behaviourally important; they are also the signal class for which our classifier was originally implemented.
The figures also decompose the F score into its components: precision and recall.
When the classifier reaches a high F score it is often achieving strong precision,
while when the PLCA does well it achieves strong recall. 

The per-class results for the most difficult evaluation condition, Cap-Field,
show that the generalisation to new individuals and new environments has a differential effect on recognition quality
(Figure \ref{fig:allrunstatsperclass}).
Importantly, the classifier-based system is able to generalise well on one of the more important categories---focal call---as well as on self-maintenance,
yet the performance on some other categories---walking, flying, bg jackdaws---drops off markedly.
The performance of the PLCA-based system does generalise on some categories---looking around, self-maintenance---but exhibits lower performance in other categories, including focal calls.

\begin{figure*}[pt]
	\centering
	{\includegraphics[page=1,width=0.9\linewidth,clip,trim=0mm 7.5mm 0mm 1mm]{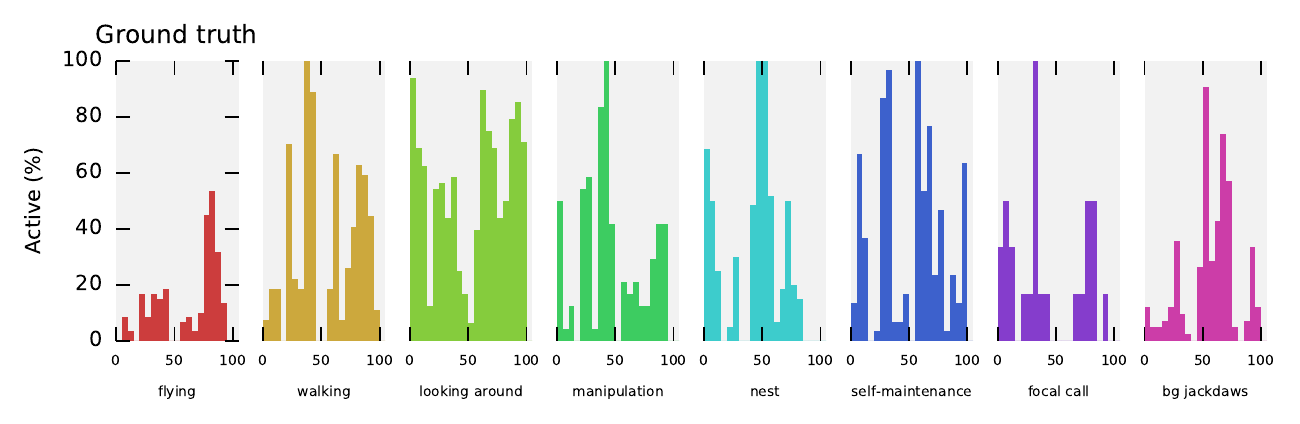}}
	{\includegraphics[page=1,width=0.9\linewidth,clip,trim=0mm 7.8mm 0mm 1mm]{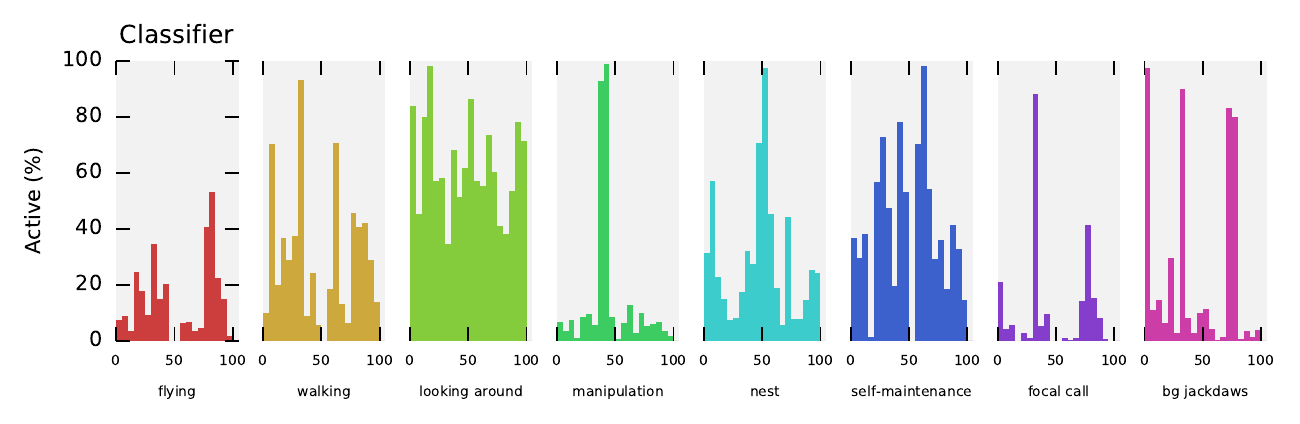}}
	{\includegraphics[page=1,width=0.9\linewidth,clip,trim=0mm 0mm 0mm 1mm]{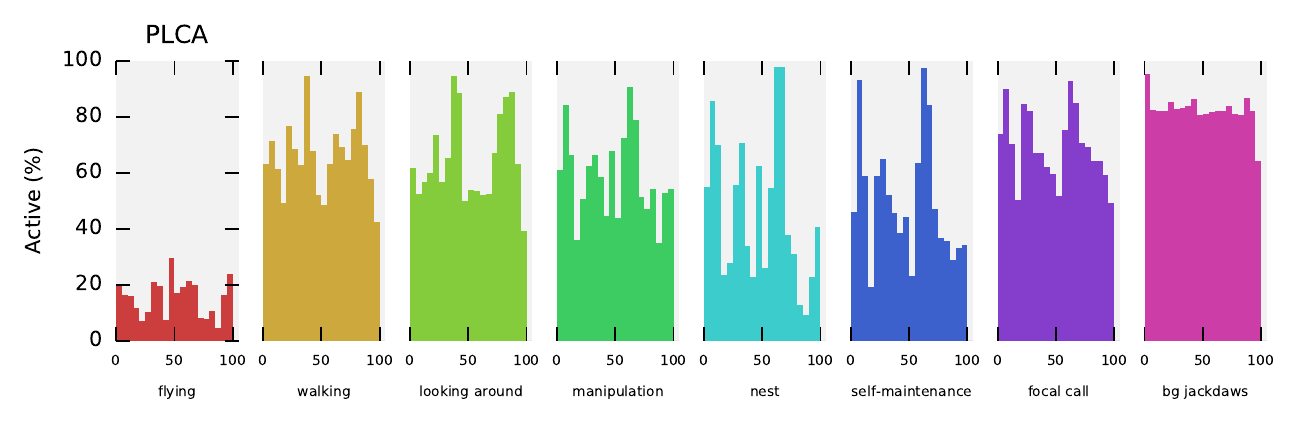}}
	\caption{Temporal activity profiles for one of the field recordings, for 8 selected classes. %
Each panel shows a bar chart plotting, for each subsequent five-minute interval, the proportion of time that the class was active. %
This was calculated as the proportion of 5-second segments in that interval that were labelled positive; %
for probabilistic outputs, the `fuzzy' probabilistic decisions were summed. %
We compare an example of the manually-annotated ground truth (top row), the classifier inference (middle row), and the PLCA inference (bottom row). The two systems were in the EachField condition, with per-class HMM filtering as postprocessing.}
	\label{fig:timebudgets}
\end{figure*}

Figure \ref{fig:timebudgets} shows a different view of the temporal nature of our data.
For selected classes in a chosen recording, it summarises the true or inferred activity levels in broad (five-minute) time-steps.
Both systems exhibit some mismatch with the ground-truth, though the output from the classifier-based system can be seen to better match the true contours of activity.
In particular the classifier-based system shows a tendency to better match the true sparsity levels of class activations.

A final comparative study was made using the higher-resolution 23~msec~step raw output of the PLCA-based system, comparing this against the 5~sec mean-pooled segments. 
Using the X-Y crossvalidation scenario, the performance in terms of segment-based F-measure with 5~sec segment size was 39.07\% when using the 23~msec output, and 38.03\% when using the 5~sec mean-pooled output. When however the high-resolution output was evaluated using the segment-based F-measure with a 100~msec segment size, performance dropped to 22.19\%. These results indicate that the higher-resolution output can lead to a small improvement over the pooled output, and that the numerical value of the chosen evaluation statistic depends strongly on the temporal granularity of evaluation.
The reduced performance when evaluated at high resolution may be partly due to issues in the temporal precision of the inferred and/or the ground-truth annotations.

\begin{figure*}[pt]
\resizebox{\linewidth}{!}{\includegraphics{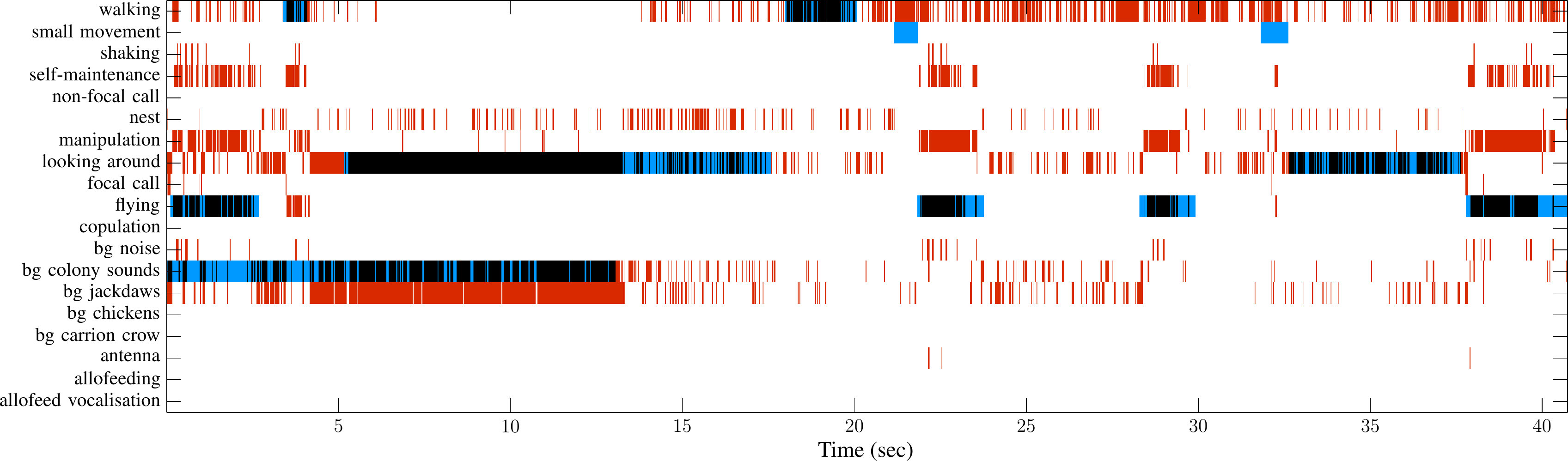}}
\caption{The 23~msec step output of a recording 
from the captive set, using the PLCA-based system with the X-Y crossvalidation scheme. The colour scheme is as in Fig. \ref{fig:wavann1}.}
\label{fig:highres}
\end{figure*}

Fig. \ref{fig:highres} shows an example high-resolution output using the PLCA-based system for recording MohawkMOV00F\_a from the captive set, which in this case reached a 100~msec segment-based F-measure of 54.1\% using the X-Y crossvaliation scheme. A few observations can be made from Fig. \ref{fig:highres}: the system was able to successfully detect overlapping contexts, in this case background colony sounds and looking around movement. However, the output was often fragmented, as for example can be seen for detected flying events. Another notable issue is the high number of false alarms as compared to missed detections (which translates into high precision and low recall, as shown in Fig. \ref{fig:allrunstatsperclass}). So for example, flight events present in the recording were correctly detected as flight, but at the same time the output produced false positives for the manipulation and self-maintenance classes.

\section{Discussion}
\label{sec:discussion}
Our study has investigated a novel task in animal sound recognition,
approaching it via two polyphonic sound recognition methodologies related to those previously studied in environmental and bird sound.
Overall evaluation figures are comparable with the state of the art in these neighbouring tasks \cite{Stowell:2015,Mesaros:2016}.
The details of the timelines recovered (Figures \ref{fig:wavann1}, \ref{fig:timebudgets}, \ref{fig:highres})
show that across all conditions, further development is needed before this paradigm can be deployed for fully automatic analysis of animal behaviour patterns from audio data.
Of the two recognition systems studied, the classifier-based system consistently led to stronger results,
including a better match to the temporal characteristics of the true annotations (Figure \ref{fig:timebudgets});
however,
the PLCA-based system has an advantage of directly outputting a high-resolution (frame-by-frame) annotation, which may be particularly desirable in some applications,
such as investigating the short-time vocal interactions between individuals.

Our sequence of crossvalidation tests demonstrated that generalising to new individuals and new environmental conditions remains a critical challenge for automatic sound recognition,
certainly when judged by F score (Figure \ref{fig:allrunstatsf}),
especially when aiming at extrapolating from captive to field datasets.
The present results suggest that to annotate field recordings, the best strategy could be to train a human annotator on the captive data to annotate a small subset of field recordings from individuals which in turn could be used to train the classifier for further field data analyses. 
Crucially, our study investigated the automatic recognition of a diverse set of classes,
each of them pertinent for the study of animal communication and behaviour.
The classes vary widely in their acoustic realisations,
from single sound events such as calls, to behaviours such as walking heard as compound events or sound textures.
Consequently, as expected there were wide variations in recognition performance across classes.
The strongest-performing system achieved good F scores for focal calls, flying, self-maintenance and walking.
In general, performance levels could be correlated with how well the class of interest was represented in the training data.
The sound of flying is quite clear to a human annotator, especially in the field where birds may fly continuously for 15 minutes or longer.
Very short flights (less than 1--2 seconds) are more difficult, and require more attention, because they may be confused e.g. with feather ruffling.
Especially the captive dataset was characterised by such short flights,
which may explain why the relatively good scores for automatic detection of flying were still lower than anticipated.
Suitable features and detectors for such noisy, loosely periodic sounds thus remain a topic for further development.

In manual inspection, we noted a tendency for systems to output detections for focal call and non-focal call at the same time.
This can be attributed partly to acoustic similarities between the classes:
the microphone placement was designed to assist with discriminating these categories,
though in some instances it remained difficult even for a human annotator.
Some acoustic differences included the effects of close-mic recording,
giving increased low-frequency energy for the focal call over the non-focal call.
We did not adapt our time-frequency representations specifically for this feature,
and one future development could include such adaptation.
A rival explanation for the confusion of focal and non-focal calls is that
the two do tend to co-occur in close temporal proximity ($<1$ seconds),
and so the systems may be influenced more by the class co-activation (at the 5-second resolution) rather than acoustics.
This highlights the tension inherent in selecting a time resolution for analysis; for studies such as this, in which the different categories operate with rather different temporal characteristics, an option may be for the system---\textit{and} also the evaluation---to use a class-dependent time resolution.

In the present study we found relatively little benefit in HMM postprocessing of system output.
Its purpose was to refine per-segment estimates by making use of temporal dependencies between segments.
In some configurations it led to a mild improvement in results,
though in some other configurations it led to deterioration. %
We did however find a consistent result that HMM filtering led to better results than Viterbi decoding,
and that a per-class HMM was better than a unified HMM.
The classifier-based system treated each segment entirely independently,
and so should have benefited from some temporal smoothing.
One interpretation is that simple Markovian dependency (at the 5-second timescale) does not reflect enough of the temporal structure present in the data,
and that more sophisticated temporal models might be investigated.

Some of the differences in interpretation implied by the AUC and the F score might be attributed to the fact that F score requires fuzzy/probabilistic outputs to be binarised at a specific threshold,
whereas the AUC uses the continuous data and thus generalises over all possible thresholds.
In a typical practical application, the user will know the relative cost of false positives and false negatives---i.e.\ the relative importance of high precision and high recall---and can set a threshold based on this balance.
The standard F score weights the two equally.
However, downstream applications might imply different priorities, such as high precision in the case of a user retrieving examples of specific behaviour.
In those cases it would be desirable to use the generalised F score, sometimes referred to as $F_\beta$ where $\beta$ is the desired precision/recall ratio. 
This would be used not only for evaluation but for threshold-setting.

As already discussed, we consider that the current level of performance is not yet at level for blind application to new data.
As with tasks in neighbouring disciplines---speaker diarisation and polyphonic music transcription---%
the task is difficult and the development of full automation will require refinement of methods adapted for the specific characteristics of the signals in question.
This is particularly true for categories indirectly represented via clusters of related sound events.
The present study with its diverse set of sound categories raises the possibility that a good detection system may benefit from using an entirely different system for each class, perhaps using different timescales.
A further possible direction in relation to the timescale is the possibility of using dynamic time resolution.
The appropriate time resolution at which to consider animal behaviour is a discussion well-rehearsed in ethology;
if time resolutions could be dynamically inferred per-class from data,
this might inform debate as well as improving system performance.

We investigated the performance of systems using segment-based evaluation measures.
Our segment size of 5 seconds was chosen based on manual inspection of pilot data as well as on considerations of the target application.
The classifier-based system was also configured to operate at this resolution;
such a classifier-based system typically operates over segments of this size
(not at `frame-wise' resolution such as 23 ms)
in order to make stable classification decisions.
Segment-based evaluations aggregate higher-resolution data using a max-pooling approach \cite{Mesaros:2016},
with the curious side-effect that a single positive item anywhere within the 5~sec segment leads to the whole segment considered active.
To mitigate this effect, in future evaluations one might use a smaller (and data-driven) segment size for evaluation, even in the case that the system gives output at a larger segment size;
perhaps more fundamentally, the max-pooling could be replaced with a parametric threshold (e.g.\ percentile-based) to reduce the effect of false-positive `blips' on the evaluation outcome.

In the present work we considered interactions between the annotated categories via co-occurrence dependencies (positive or negative) implicitly learnt from the data:
the classifier-based system used a single classifier predicting for all classes at once,
the PLCA-based system had the opportunity to `explain away' a portion of energy as belonging to one class rather than another,
and the HMM postprocessing was able to use a single HMM model across all classes (though this was not found to be better than per-class HMMs).
Future work could consider alternative approaches to the relationships between categories.
Hierarchical models such as the context-dependent sound event detection of \cite{Heittola13}
may be suitable, or switching state-space models (switching SSMs),
where the discrete ``switch'' would correspond to a context and the context-dependent SSMs would detect specific sound events or background sounds.

\section{Conclusions}
\label{sec:conc}
We have introduced an application of audio recognition specifically for sound recordings from animal-attached microphones, to enable analysis of the activity of a focal animal as well as the context of such activity, i.e. the environment around it as conveyed acoustically.
This enables researchers to study the animal's behaviour as well as the context of that behaviour, i.e. the environment around it as conveyed acoustically.
We applied automatic recognition to data collected from lightweight backpack loggers carried by free-flying birds (jackdaws) in an aviary and in the field.

We directly compared a scene-classification and an event-detection approach approach to this task.
The classification method made use of a feature learning method developed for bird vocalisations.
For event detection, we introduced a modified PLCA method, improving on previously-published work in related domains.
In evaluation, the classifier-based method performed most strongly.

We find that the current recognition quality level enables scalable automatic annotation of audio logger data, given partial annotation, but also find that individual differences between animals and/or their backpacks can reduce recognition rates when generalising to previously-unseen individuals.
This approach to studying animal behaviour in single individuals requires further development for full automation and application to previously-unseen individuals.
However, as on-animal microphones become increasingly common, this seems an effort worth taking to eventually extract meaning from such streams of sounds by facilitating the analyses of vocalisations, as well as some of their associated behaviours and acoustic contexts, without additional data collection and devices. Combining such results with an animal's position in space or relative to its conspecifics, and with detailed acceleration data, would provide us with a more complete picture of what animals do and even provide hints why they do it, to tackle many remaining open questions in mechanistic, evolutionary and conservation-related areas of behavioural research.


\section*{Author Contributions}

DS and LFG jointly conceived the study.
DS implemented the classifier-based system, led on the evaluation, led on the manuscript writing and wrote parts of the manuscript.
LFG 
 provided the data (conducted animal handling and fieldwork, supervised video recordings, performed all initial sound and video analyses and annotations), helped with evaluating the method, and wrote parts of the manuscript.
EB implemented the PLCA-based system, collaborated in performing the evaluation, and wrote parts of the manuscript.

\section*{Acknowledgments}

We would like to thank Tiffany Magdalena Pelayo van Buuren and Magdalena Mair for assistance in the field; Katrin Mayer for collecting video footage and helping with sound files; Auguste von Bayern, Wolfgang Goymann, Andries Ter Maat and Manfred Gahr for their support without which this work had not been possible; and to the von Bayern family for granting access to the premises and facilities.

DS is supported by EPSRC Early Career research fellowship EP/L020505/1.
EB is supported by a UK Royal Academy of Engineering Research Fellowship (grant no. RF/128).
LFG is funded by the Max Planck Society.

\bibliographystyle{IEEEtran}
\bibliography{jackdawcontext}

\end{document}